\documentclass[
nobibnotes,aps,pre,showkeys,
12pt]{revtex4-2}

\usepackage{amssymb}
\usepackage{amsmath}
\usepackage{color}

\begin{document}




\title{Two-dimensional Fireballs as a Lagrangian Ermakov System}


\author{Fernando Haas} 

\affiliation{Physics Institute, Federal University of Rio Grande do Sul, 91501-970, Av. Bento Goncalves 9500, Porto Alegre, RS, Brazil}

\begin{abstract}
The equations of motion for the variance of strictly one-dimensional or two-dimensional non-relativistic fireballs are derived, from the hydrodynamic equations for an ideal, structureless Boltzmann gas. For this purpose a Gaussian number density {\it Ansatz} is applied, together with low-dimensional proposals for the energy density, coherent with the equipartition theorem. The resulting ordinary differential equations are shown to admit a variational formulation. The underlying symmetries are connected to constants of motion, through Noether's theorem. The two-dimensional case is special, corresponding to a Lagrangian Ermakov system without external forcing. There is a comparison with the fully three-dimensional fireballs, and its reduction to effective two-dimensional dynamical system for elliptic trajectories. The exact analytical solutions are worked out.
\end{abstract}









\maketitle

\section{Introduction}

The dynamics of expanding fireballs has recently attracted renewed attention, providing an instance where the hydrodynamic partial differential equations can be reduced to a set of ordinary differential equations. As analyzed in recent works,  through suitable equations of state and Gaussian density proposals the reduced models are exactly integrable, including rotation and multi-particle effects in non-relativistic or relativistic flows \cite{r1, r2, r3, r4}. Applications are related to quark matter and strongly interacting hadronic matter \cite{r1, r4}, in non-relativistic or relativistic heavy ions collisions \cite{r5, r6}. 

The present work provides an alternative view on the reasons underlying the existence of these exact solutions, in some situations. Namely, the presence of a scaling symmetry is explored, yielding a connection between symmetry and conservation laws. This is developed in view of the underlying variational formulation, by means of Noether's theorem \cite{Noether, Sarlet}. Moreover, the case of two-dimensional (2D) flows is shown to be reducible to Ermakov systems \cite{Ermakov, Ray}. The Ermakov constant of motion is then verified to be a Noether invariant, due to a dynamical symmetry of the action functional. Completely integrable Ermakov systems with a variational formulation (Lagrangian, Hamiltonian) have been originally discussed in \cite{c1, c2}. Noether symmetries for the Ermakov system and related systems involving the associated Pinney equation \cite{pin} have found applications in 2D Bose-Einstein condensates \cite{c3}, graphene gauged bilayer models \cite{gra}, Reid systems \cite{manc}, integrable equations with Ermakov–Pinney nonlinearities and Chiellini damping \cite{manca}, two-dimensional affine subalgebras of Ermakov–Pinney Lie algebra \cite{cari}, parametric oscillators \cite{gale} and Master-Slave systems of damped Duffing oscillators \cite{barba}. A recent review on Ermakov systems and their related symmetries can be found in \cite{cerv}. The dynamics of fireballs has peculiar properties according to the number of spatial dimensions \cite{r1, r2, r3, r4}. This is similar to stellar structure considering N-dimensional white dwarf stars \cite{Chavanis}. 

The article is organized as follows.  In Section 2, the equations of motion for an ideal, structureless Boltzmann fluid 2D fireball are put forward.  Section 3 rewrites the equations of motion as a Lagrangian Ermakov system. The point and dynamical symmetries together with the constants of motion are worked out, as well as the exact analytical solution. Section 4 provides a comparison with 3D fireballs, also showing elliptic flows as a Lagrangian Ermakov system. For completeness, the 1D fireball dynamics is analyzed in Section 5. Section 6 is devoted to a final discussion.   

\section{Two-dimensional Fireballs}

Here we provide the adaptation of the fireballs model from \cite{r1, r2} to a strictly 2D fluid. To begin with, consider the well-known hydrodynamic equations 
\begin{eqnarray}
\label{e1}
\frac{\partial n}{\partial t} + \nabla\cdot(n {\bf v}) &=& 0 \,,\\
\label{e2}
\left(\frac{\partial}{\partial t} + {\bf v}\cdot\nabla\right) {\bf v} &=& - \frac{\nabla p}{m n} \,,\\
\label{e3}
\frac{\partial \varepsilon}{\partial t} + \nabla\cdot(\varepsilon {\bf v}) &=& - p \nabla\cdot{\bf v} \,,
\end{eqnarray}
where $n = n({\bf r}, t)$ is the particle number density, ${\bf v} = {\bf v}({\bf r}, t) = (v_x, v_y)$ is the fluid velocity field, $p = p({\bf r}, t)$ is the pressure, $m$ is the particle mass and $\varepsilon = \varepsilon({\bf r}, t)$ is the energy density, with ${\bf r} = (x, y)$. For closure, for simplicity it is assumed the equations of state for an ideal (collisionless) , structureless Boltzmann gas, namely
\begin{equation}
\label{eos}
p = n T \,, \quad \varepsilon = n T \,,   
\end{equation}
where $T = T({\bf r}, t)$ is the local temperature. Notice the second member in Eq. (\ref{eos}) follows from equipartition theorem $\varepsilon = (D/2) n T$ where $D = 2$ is the number of degrees of freedom. Units with a Boltzmann constant $\kappa_B = 1$ will be used throughout. 

In passing, notice the constancy of the total energy ${\cal H}$ or $d{\cal H}/dt = 0$, where
\begin{equation}
\label{te}
{\cal H} = \int dx dy \left(\frac{n m v^2}{2} + \varepsilon\right) \,, \quad v^2 = v_x^2 + v_y^2 \,.
\end{equation}

To proceed, it is possible to convert the system of Eqs. (\ref{e1}-\ref{e3}) supplemented by Eq. (\ref{eos}) to a set of ordinary differential equations, using the very same {\it Ansatz} of \cite{r1, r2, r3, r4} adapted to a fluid constrained to a plane, namely 
\begin{eqnarray}
\label{n1}
n &=& n_0 \left(\frac{X_0 Y_0}{X Y}\right)\,\exp\left(- \frac{x^2}{2 X^2} - \frac{y^2}{2 Y^2}\right)  \,,\\
v_x &=& \frac{\dot X}{X}\,x \,, \quad v_y = \frac{\dot Y}{Y}\,y \,, \\
\label{n3}
T &=& T_0 \left(\frac{X_0 Y_0}{X Y}\right) \,,
\end{eqnarray}
where $X = X(t), Y = Y(t)$ are functions of time only, and $n_0, T_0, X_0, Y_0$ are positive reference values. In the 3D case with variances $X, Y, Z$ one has \cite{r1, r2, r3, r4} an {\it Ansatz} with factor $1/(X Y Z)^{2/3}$ - which is also an inverse quadratic form - instead of the factor $1/(X Y)$ as in Eqs. (\ref{n1}) and (\ref{n3}). In particular, this is due to a proper normalization of the number density, and to complain with the continuity equation. As a simple algebra shows, the ultimate reason for the proposed Eqs. \ref{n1}-\ref{n3} is that it reduces the fluid equations to a set of ordinary differential equations, in the same trend of the 3D case\cite{r1, r2, r3, r4}, now constrained to two spatial dimensions. 

The resulting system, as can be verified after a trivial calculation, reads
\begin{equation}
\label{e4}
\ddot X = \frac{T_0}{m X} \left(\frac{X_0 Y_0}{X Y}\right) \,, \quad \ddot Y = \frac{T_0}{m Y} \left(\frac{X_0 Y_0}{X Y}\right) \,,
\end{equation}
where a dot means derivative with respect to the time. 

In the following, we apply the rescaling
\begin{equation}
\label{ee}
\bar X = \frac{X}{\sqrt{X_0 Y_0}} \,, \quad \bar Y = \frac{Y}{\sqrt{X_0 Y_0}} \,, \quad \bar t = \left(\frac{T_0}{m X_0 Y_0}\right)^{1/2}\,t \,.   
\end{equation}
Omitting bars, from Eqs. (\ref{e4})-(\ref{ee}) the resulting system in terms of the dimensionless variables the system reads  
\begin{equation}
\label{e5}
\ddot X = \frac{1}{X^2 Y}  \,, \quad \ddot Y = \frac{1}{X Y^2}  \,,
\end{equation}
whose properties will be discussed in what follows.

It should be noted, that unlike the previous literature, here the equation of state for the energy density is adapted to a true 2D fluid, with two degrees of freedom only. One could, for instance, start with a fluid in three spatial dimensions, with $\varepsilon = (3/2) n T$ as in Eq. (4) of \cite{r1} and then reduce to specialized solutions e.g. elliptic flows. This would lead to a pair of second-order ordinary differential equations, as in an effectively 2D mechanical system, but within a 3D context.

\section{Analytical Solutions, Symmetries and Invariants}

It happens that the pair in Eq. (\ref{e5}) is an autonomous, Hamiltonian Ermakov system, with a pseudo-potential $V$, 
\begin{equation}
\label{e6}
\ddot X = - \frac{\partial V}{\partial X} \,, \quad \ddot Y = - \frac{\partial V}{\partial Y} \,, \quad V = V(X, Y) = \frac{1}{X Y} \,.
\end{equation}
In this situation, the equations of motion can be exactly solved.

In the most usual version \cite{Ray}, Ermakov systems are given  by 
\begin{eqnarray}
\label{ermak}
\ddot X + \omega^2(t) X = \frac{f(Y/X)}{X^2 Y}  \,, \quad \ddot Y + \omega^2(t) Y = \frac{g(X/Y)}{X Y^2} \,,
\end{eqnarray}
where $\omega = \omega(t)$ is a time-dependent frequency and $f = f(Y/X)$ and $g = g(X/Y)$ are generic functions of the indicated arguments.  
Ermakov systems always admit the invariant 
\begin{equation}
\label{RR}
I = \frac{1}{2} (X \dot Y - Y \dot X)^2 + \int^{Y/X} f(s) ds +  \int^{X/Y} g(s) ds   \,,
\end{equation}
which is a constant of motion, $dI/dt = 0$ along trajectories.

To cast Eq. (\ref{e6}) into Ermakov form, it is only necessary to choose $\omega = 0, f = g =1$. The choice $\omega = 0$ is due to the absence of any external force in the hydrodynamic model. In particular, the Ermakov form would be also achieved with $\omega \neq 0$, in the case of a time-dependent harmonic external trap.  The corresponding Ermakov invariant is
\begin{equation}
\label{erma}
I = \frac{1}{2} (X \dot Y - Y \dot X)^2 + \frac{Y}{X} + \frac{X}{Y}   \,.
\end{equation}
Moreover, it is immediate to write the energy first integral as
\begin{equation}
\label{h}
H = \frac{1}{2}(\dot X^2 + \dot Y^2) + \frac{1}{X Y} \,,   
\end{equation}
such that $dH/dt = 0$ along trajectories. It is possible to show that $H$ exactly corresponds to the total fluid energy in Eq. (\ref{te}), in terms of rescaled variables.

Being a 2D autonomous Hamiltonian system admitting the two involutive constants of motion $H, I$, it would be tempting to identify the system in Eq. (\ref{e6}) as completely integrable. However, strictly speaking the first integrals need to have compact level surfaces in phase space, for complete integrability \cite{Arnold}, allowing the introduction of action-angle variables. This is not presently the case. Nevertheless, the equations of motion are reducible to quadrature, as shown in the following.

For the exact solution, introduce polar coordinates $(r, \phi)$ so that $X = r \cos\phi, Y = r \sin\phi$. The energy first integral 
becomes 
\begin{equation}
\label{hh}
H = \frac{\dot r^2}{2} + \frac{I}{r^2} \,,
\end{equation}
with Ermakov invariant 
\begin{equation}
\label{ei}
I = \frac{1}{2}(r^2 \dot\phi)^2 + \frac{1}{\sin\phi \cos\phi}    \,.
\end{equation}

Assuming for simplicity a large enough initial angular momentum to assure $I > 0, H > 0$, it is straightforward to integrate Eq. (\ref{hh}) once again, to find 
\begin{equation}
\label{rt}
r = \left(2 H (t - t_0)^2 + I/H\right)^{1/2} \,, \quad t_0 = {\rm cte} \,.    
\end{equation}
By inspection, for large enough times one has a ballistic expansion $r \sim t$. 

Finally, for the angular variable it is useful to introduce a new time variable 
\begin{equation}
\label{ntv}
\tilde{t} = \int_{t_0}^t  \frac{dt'}{r^{2}(t')} =  \frac{1}{\sqrt{2 I}} \,{\rm arctan}\left(\sqrt{\frac{2}{I}} H (t - t_0)\right) \,,  
\end{equation}
converting the Ermakov invariant into an energy-like form,
\begin{equation}
\label{el}
I = \frac{1}{2}\left(\frac{d\phi}{d\tilde{t}}\right)^2 + \frac{1}{\sin\phi \cos\phi}    \,.
\end{equation}
Equation (\ref{el}) can be integrated by quadrature in terms of elliptic functions, but the exact solution is an awkward expression, omitted here.   


It is possible to identity that the system in Eq. (\ref{e5}) is invariant under a scaling transformation, with generator 
\begin{equation}
\label{g}
G = 2 t \frac{\partial}{\partial t} + X \frac{\partial}{\partial X} + Y \frac{\partial}{\partial Y}  \,,
\end{equation}
as shown in the Appendix.
The generator of the first extended group is 
\begin{equation}
G^{[1]} = G - \dot{X} \frac{\partial}{\partial\dot X} - \dot Y \frac{\partial}{\partial\dot Y}  \,.
\end{equation}

It is easy to verify that the Ermakov constant of motion is invariant under the scaling symmetry
\begin{equation}
 G^{[1]}\,I = 0 \,.   
\end{equation}
In this context, $I$ can be viewed as a consequence of the scaling symmetry. 

Moreover, the scaling transformation from Eq. (\ref{g}) can be shown to be a Noether symmetry. According to Noether's theorem \cite{Noether, Sarlet}, the invariance of the action functional (up to addition of a numerical constant) can be associated to a constant of motion. 
For a Lagrangian $L = L({\bf R},\dot{\bf R},t)$, where ${\bf R} = (X, Y)$, consider a general point symmetry generator 
\begin{equation}
G = \tau \frac{\partial}{\partial t} + \boldsymbol{\eta} \cdot\frac{\partial}{\partial{\bf R}} \,,    
\end{equation}
where $\tau = \tau({\bf R}, t), \, \boldsymbol{\eta} = \boldsymbol{\eta}({\bf R}, t) = (\eta_X, \eta_Y)$. The corresponding  generator of the first extended group is then 
\begin{equation}
G^{[1]} = G + (\dot{\boldsymbol{\eta}} - \dot\tau \dot{\bf R})\cdot\frac{\partial}{\partial\dot{\bf R}} \,,    
\end{equation}
where 
\begin{equation*}
\dot\tau = \frac{\partial\tau}{\partial t} + \dot{\bf R}\cdot\frac{\partial \tau}{\partial {\bf R}} \,,  \quad \dot{\boldsymbol{\eta}} = \frac{\partial\boldsymbol{\eta}}{\partial t} + \dot{\bf R}\cdot\frac{\partial \boldsymbol{\eta}}{\partial {\bf R}} \,.
\end{equation*}
The Noether symmetry condition is
\begin{equation}
\label{nsc}
G^{[1]}L + \dot\tau L = \dot\Lambda    \,, \quad \dot\Lambda = \frac{\partial\Lambda}{\partial t} + \dot{\bf R}\cdot\frac{\partial \Lambda}{\partial {\bf R}} \,,
\end{equation}
where the gauge function $\Lambda = \Lambda({\bf R},t)$ for a point transformation. The associated Noether invariant is 
\begin{equation}
\label{ni}
J = \tau \left(\dot{\bf R}\cdot\frac{\partial L}{\partial\dot{\bf R}} - L\right)  -   \boldsymbol{\eta}\cdot\frac{\partial L}{\partial\dot{\bf R}} + \Lambda \,, \quad \frac{dJ}{dt} = 0 \,. 
\end{equation}
As usual, the energy first integral in Eq. (\ref{h}) comes from time translation invariance, with $\tau = 1, \boldsymbol{\eta} = 0, \Lambda = 0$. 

The 2D fireball system in Eq. (\ref{e5}) possess the Lagrangian 
\begin{equation}
\label{l}
L = \frac{1}{2}(\dot X^2 + \dot Y^2) - \frac{1}{X Y} \,,   
\end{equation}
which is the Legendre transform of $H$ in Eq. (\ref{h}). 

It happens that the Noether symmetry condition (\ref{nsc}) is satisfied for $G$ in Eq. (\ref{g}) and $L$ in Eq. (\ref{l}), taking $\Lambda = 0$. The corresponding Noether invariant from Eq. (\ref{ni}) is the explicitly time-dependent constant of motion 
\begin{equation}
J  = 2 t  \left(\frac{1}{2}(\dot X^2 + \dot Y^2) + \frac{1}{X Y}\right) - X \dot X - Y \dot Y \,. 
\end{equation}
This is therefore a third independent constant of motion for the 2D fireball system. 

Although the Noether invariant from the scaling generator in Eq. (\ref{g}) is not the Ermakov invariant, it happens that $I$ results from a dynamical Noether symmetry given by 
\begin{eqnarray}
\nonumber 
\bar{X} &=& X + \epsilon [\tau\dot X + Y (X \dot Y - Y \dot X)] \,,    \\
\bar{Y} &=& Y + \epsilon [(\tau\dot Y - X (X \dot Y - Y \dot X)] \,,    \\
\nonumber
\bar{t} &=& t + \epsilon \tau \,,
\end{eqnarray}
where $\epsilon$ is an infinitesimal parameter and $\tau = \tau({\bf R}, \dot{\bf R}, t)$ is an arbitrary function of the indicated arguments. Notice the intrinsic dependence of the transformation on the velocity. The associated generator of dynamical symmetry is 
\begin{equation}
\label{ds}
G = \tau \frac{\partial}{\partial t} + [\tau\dot X + Y (X \dot Y - Y \dot X)] \frac{\partial}{\partial X} + 
+ [\tau\dot Y - X (X \dot Y - Y \dot X)] \frac{\partial}{\partial Y}\,.    
\end{equation}

The Noether symmetry condition (\ref{nsc}) is satisfied with the $G$ from Eq. (\ref{ds}), with the gauge function
\begin{equation}
 \Lambda = \tau L - \frac{1}{2}(X \dot Y - Y \dot X)^2 + \frac{X}{Y} + \frac{Y}{X} \,.
\end{equation}
Notice now $\Lambda$ is intrinsically dependent on the velocities too. 
The associated Noether invariant from Eq. (\ref{ni}) coincides with the Ermakov invariant in Eq. (\ref{erma}), which at the end is $\tau$-independent. In this context, the Ermakov invariant exists due to a dynamical Noether symmetry. For Lagrangian Ermakov systems in general, the Ermakov invariant can be viewed as a Noether invariant with dynamical symmetry, as found from the converse of Noether's theorem \cite{hg}.

\section{Comparison with 3D Fireballs}

The 3D fireball system derived in \cite{r1} is given by
\begin{equation}
\label{ee5}
\ddot X = \frac{1}{X (X Y Z)^{2/3}}  \,, \quad \ddot Y = \frac{1}{Y (X Y Z)^{2/3}}  \,, \quad \ddot Z = \frac{1}{Z (X Y Z)^{2/3}} \,,
\end{equation}
in terms of dimensionless variables. It has the energy first integral 
\begin{equation}
\label{hhh}
H = \frac{1}{2}(\dot X^2 + \dot Y^2 + \dot Z^2) + \frac{3}{2} (X Y Z)^{- 2/3} \,,   
\end{equation}
corresponding to the conserved total fluid energy, in terms of rescaled variables, 
together with a Lagrangian 
\begin{equation}
\label{ll}
L = \frac{1}{2}(\dot X^2 + \dot Y^2 + \dot Z^2) - \frac{3}{2} (X Y Z)^{- 2/3} \,.   
\end{equation}
The interpretation of $X, Y, Z$ is similar to the 2D case, namely, it corresponds to variances of a Gaussian number density in each respective spatial direction \cite{r1}. 

Now we show that it happens that Eq. (\ref{ee5}) also possess a scaling symmetry, with generator 
\begin{equation}
\label{gg}
G = 2 t \frac{\partial}{\partial t} + X \frac{\partial}{\partial X} + Y \frac{\partial}{\partial Y} + Z \frac{\partial}{\partial Z}  \,.
\end{equation}
The generator of the first extended group is 
\begin{equation}
G^{[1]} = G - \left(\dot{X} \frac{\partial}{\partial\dot X} + \dot Y \frac{\partial}{\partial\dot Y} + \dot Z \frac{\partial}{\partial\dot Z}\right)  \,.
\end{equation}
Similarly to the 2D case, the scaling is also a Noether symmetry, since 
\begin{equation}
\label{nnsc}
G^{[1]}L + \dot\tau L = 0 \,. 
\end{equation}
The corresponding Noether invariant from Eq. (\ref{ni}) easily adapted to three spatial dimensions, is
\begin{equation}
\label{RRR}
J = 2 t H - {\bf R}\cdot\dot{\bf R} \,, \quad \frac{dJ}{dt} = 0 \,,     
\end{equation}
where $H$ is given by Eq. (\ref{hhh}) and $ {\bf R} = (X, Y, Z)$. 

Interestingly, in terms of spherical coordinates
\begin{equation}
 X = r\cos\phi\sin\theta \,, \quad Y = r\sin\phi\sin\theta \,, \quad Z = r\cos\theta 
 \end{equation}
one has $J = 2 t H - r\dot r$, which can be integrated again yielding
\begin{equation}
r^2 = 2 H t^2 - 2 J t + r_0^2 \,, \quad r_0 = r(0) \,.
\end{equation}
This shows indeed an exploding fireball for appropriate parameters.

\subsubsection{Elliptic Fireballs: a 2D Lagrangian Ermakov System}

Following \cite{r1}, it is possible to consider elliptic solutions such that $Z = X$, yielding from Eq. (\ref{ee5}),
\begin{equation}
\label{ee55}
\ddot X = \frac{1}{X (X^2 Y)^{2/3}}  \,, \quad \ddot Y = \frac{1}{Y (X^2 Y)^{2/3}}  \,.
\end{equation}
The exact solution was found in \cite{r1}, in terms of stretched radial coordinates. Here it is shown, that the integration of the system (\ref{ee55}), comes from expressing it as a 2D Lagrangian Ermakov system admitting the energy first integral and the Ermakov invariant.  

The Ermakov form in Eq. (\ref{ermak}) is achieved choosing 
\begin{equation}
 f\left(\frac{Y}{X}\right) = g\left(\frac{X}{Y}\right) = \left(\frac{Y}{X}\right)^{1/3} \,, \quad \omega(t) = 0 \,,   
\end{equation}
yielding the Ermakov invariant 
\begin{equation}
\label{iii}
I = \frac{1}{2}(X \dot Y - Y \dot X)^2 + \frac{3}{4}\left(\frac{Y}{X}\right)^{4/3} + \frac{3}{2}\left(\frac{X}{Y}\right)^{2/3} \,, \quad \frac{dI}{dt} = 0 \,.
\end{equation}
In the sequence it is useful to consider $\tilde I = 2 I$. 

A Lagrangian function for the system in Eq. (\ref{ee55}) is 
\begin{equation}
\label{ss}
 L = \dot{X}^2 + \frac{\dot Y^2}{2} - \frac{3}{2} (X^2 Y)^{- 2/3} \,.  
\end{equation}
which comes from Eq. (\ref{ll}) upon inserting $Z = X$. As can be directly verified, the Euler-Lagrange from Eq. (\ref{ss}) gives the system (\ref{ee55}) for elliptic trajectories. 

Moreover, the system has the energy first integral 
\begin{equation}
\label{ii}
H = \dot{X}^2 + \frac{\dot Y^2}{2}  + \frac{3}{2} (X^2 Y)^{- 2/3}  \,.
\end{equation}
This can be found from the Legendre transform (with canonical momenta $(P_X, P_Y) = (2\dot X, \dot Y))$.

The Lagrangian Ermakov formulation, provides a deeper insight, about why the elliptic fireball system is exactly soluble. Following \cite{r1}, consider stretched polar coordinates, so that 
\begin{equation}
X = \frac{1}{\sqrt{2}} r \cos\phi \,, \quad Y = r \sin\phi \,,      
\end{equation}
in terms of which 
\begin{eqnarray}
\label{jjj}
 \tilde I &=& \frac{1}{2} \left(r^2 \dot\phi\right)^2 + \frac{3}{2^{1/3}} (\cos^{2}\phi \sin\phi)^{-2/3} \,, \\   
 H &=& \frac{\dot r^2}{2} + \frac{\tilde I}{r^2} \,.
\end{eqnarray}
Now the conserved energy is exactly the same as in Eq. (\ref{hh}), so that the radial variable is also given by Eq. (\ref{rt}), but with the replacement $I \rightarrow \tilde I$, assuming $\tilde I > 0$ for simplicity.

For completeness, we write the Noether invariant for elliptic fireballs, namely 
\begin{equation}
J = 2 t H - (2 X \dot X + Y \dot Y) \,, \quad \frac{dJ}{dt} = 0 \,,  
\end{equation}
an expression directly arising from (\ref{RRR}) after setting $Z = Y$. It can be easily shown that the Noether invariant in this case is also a result from a scaling transformation as a dynamical Noether
symmetry.

In the same trend, introducing the new time variable $\tilde t$ from Eq. (\ref{ntv}) converts the Ermakov invariant into the energy-like form 
\begin{equation}
\label{elq}
\tilde I = \frac{1}{2}\left(\frac{d\phi}{d\tilde{t}}\right)^2 + \frac{3}{2}(\cos\phi^2 \sin\phi)^{-2/3}    \,.
\end{equation}
which is not quite the same expression as for true 2D fireballs, shown in Eq. (\ref{el}) \,.
Nevertheless, it is possible to further integrate Eq. (\ref{elq}) to again obtain the angular variable in terms of an awkward  expression involving elliptic functions \cite{r1}.

It can be observed, that the system (\ref{e5}) describes a true 2D fireball, with fluid flow restricted to two degrees of freedom, while the system (\ref{ee55}) comes from special, elliptic trajectories after setting $Z = Y$ in the context of  3D fireball dynamics.

\section{1D Fireballs}

For completeness, it is interesting to briefly examine the case of a fluid constrained to move in an 1D (one-dimensional) channel. The hydrodynamic equations are then 
\begin{eqnarray}
\label{ee1}
\frac{\partial n}{\partial t} + \frac{\partial}{\partial x}(n v) &=& 0 \,,\\
\label{ee2}
\frac{\partial v}{\partial t} + v \frac{\partial v}{\partial x} &=& - \frac{1}{m n} \frac{\partial p}{\partial x} \,,\\
\label{ee3}
\frac{\partial \varepsilon}{\partial t} + \frac{\partial}{\partial x}(\varepsilon v) &=& - p \frac{\partial v}{\partial x} \,,
\end{eqnarray}
where $n = n(x,t)$ is the 1D particle number density, $v = v(x,t)$ is the 1D local velocity, $\varepsilon = \varepsilon(x,t)$ is the 1D local energy density, $p = p(x,t)$ is the pressure, $m$ is the particles mass. 

In exact correspondence with the derivation of the 2D fireball system, we assume for simplicity the equations of state for an  
ideal, structureless Boltzmann gas, 
\begin{equation}
\label{eeos}
p = n T \,, \quad \varepsilon = \frac{n T}{2} \,,    
\end{equation}
where $T = T(x,t)$ is the local temperature. The energy density expression, follows from the equipartition theorem for a system with one degree of freedom. 

To map the hydrodynamic equations to an ordinary differential equation, suppose 
\begin{equation}
n = n_0 \left(\frac{X_0}{X}\right)\,\exp\left(- \frac{x^2}{2 X^2}\right)  \,, \quad 
v = \frac{\dot X}{X}\,x \,, \quad T = T_0 \left(\frac{X_0}{X}\right)^2 \,,
\end{equation}
where $X = X(t)$ is a function of time only, and $n_0, T_0, X_0$ are positive reference values. 

The result is 
\begin{equation}
\ddot X = \frac{T_0}{m X} \left(\frac{X_0}{X}\right)^2 \,,    
\end{equation}
which can be put into dimensionless form using 
\begin{equation}
\bar{X} = \frac{X}{X_0} \,, \quad \bar{t} = \left(\frac{T_0}{m}\right)^{1/2} \frac{t}{X_0} \,,    
\end{equation}
yielding 
\begin{equation}
\label{1d}
\ddot X = \frac{1}{X^3} \,,   
\end{equation}
omitting the bars for brevity. Equation (\ref{1d}) is a Pinney equation \cite{pin} with zero frequency. 

Equation (\ref{1d}) is readily solved, 
\begin{equation}
\label{xxx}
X = \left(2 H (t - t_0)^2 + \frac{1}{2 H}\right)^{1/2} \,, \quad t_0 = {\rm cte.} \,,     
\end{equation}
where 
\begin{equation}
\label{h1d}
H = \frac{\dot X^2}{2} + \frac{1}{2 X^2}     
\end{equation}
is the conserved energy first integral. Like for 2D and 3D, $H$ comes from the conservation of the total fluid energy,
in terms of rescaled variables. 

In complete correspondence with the 2D and 3D cases, the scaling symmetry with generator 
\begin{equation}
G = 2 t \frac{\partial}{\partial t} + X \frac{\partial}{\partial X}    
\end{equation}
is a Noether symmetry, in a description with the Lagrangian
\begin{equation}
L = \frac{1}{2}\dot X^2 - \frac{1}{2 X^2} \,.
\end{equation}
The Noether invariant is 
\begin{equation}
J = 2 t H - X \dot X \,,  \quad  \frac{dJ}{dt} = 0 \,,  
\end{equation}
where $H$ is given by Eq. (\ref{h1d}).

Although the equation (\ref{1d}) is trivial, it still can be viewed as a member of the Ermakov pair shown in Eq. (\ref{ermak}), where $\omega = 0$,  $f(Y/X) = Y/X$ and $g(X/Y)$ is an arbitrary function of the indicated argument. As an example, one can freely chose $g = 0$, so that 
\begin{equation}
\ddot Y = 0 \,,
\end{equation}
and from Eq. (\ref{RR}) we have 
\begin{equation}
\label{11dd}
I = \frac{1}{2}(X \dot Y - Y \dot X)^2 + \frac{1}{2}\frac{Y^2}{X^2} \,.
\end{equation}

Even if not mandatory in this simple case, one might somehow interpret the invariant (\ref{11dd}) following the usual quadrature procedure by means of a nonlinear superposition law \cite{reid}, introducing variables 
\begin{equation}
\label{yyyy}
 u = \frac{Y}{X} \,, \quad \tau = \int \frac{dt}{X^2} \,,
\end{equation}
so that 
\begin{equation}
I = \frac{1}{2}\left(\frac{du}{d\tau}\right)^2 + \frac{u^2}{2} \,,
\end{equation}
which can be integrated to 
\begin{equation}
\label{uu}
u = \sqrt{2I}\sin\tau \,,
\end{equation}
skipping an integration constant to have nicer expressions at the end. 
One can the have $Y$ from Eq. (\ref{yyyy}) once a particular solution $X$ is found.  We choose 
\begin{equation}
X = \left[(t - t_0)^2 + 1\right]^{1/2} \,,
\end{equation}
found from Eq. (\ref{xxx}) with $H = 1/2$. In this case, $\tau = \arctan(t - t_0)$ and finally 
\begin{equation}
Y = \sqrt{2I}(t - t_0) \,,
\end{equation}
solving the free particle equation in terms of the Ermakov invariant as an integration constant. Such an avenue is possible due to the quite degenerate case $\omega = 0$. Obviously more complicated choices for the arbitrary function $g(X/Y)$ are also allowable.


\section{Conclusions}

We have detailed the variational formulation of fireball dynamics in 1D, 2D and 3D cases. The resulting conventional Ermakov form is a peculiar property of 2D fireballs, be it for strictly two degrees of freedom (Section 3) or for elliptic flows evolving in a 3D space (Section 4). 
The final equations of motion are always second-order, autonomous ordinary differential equations with effective inverse cubic force, as apparent form Eqs. (\ref{e5}), (\ref{ee5}), (\ref{ee55}) and (\ref{1d}). The associated scaling symmetry is also a Noether symmetry in these cases.  The Lagrangian Ermakov cases have an additional Noether dynamical symmetry, with Ermakov invariant as the Noether invariant. The exact solutions can be found thanks to the existence of the the first integrals, together with appropriate change of variables whenever necessary. The results are of potential application for fireball hydrodynamics in general, as in strongly interacting hadronic or quark matter, taking into account the necessary adaptations  (e.g. relativistic hydrodynamics). In a general context, the derivation of  analytical solutions, variational formulations, symmetries and exact conservation laws, are relevant topics in fluid mechanics.  Although the original hydrodynamic models have been taken as simple enough (collisionless, structureless, no external force) the chosen methods can inspire extensions to more general systems. 

{\vskip .5cm}
{\bf Appendix: derivation of the symmetry generator in Eq. (\ref{g}) for the system in Eq. (\ref{e5}).}
{\vskip .5cm}

The form of the system in Eq. (\ref{e5}) makes us suspect that it admits a scaling symmetry in terms of new variables $\bar{X}, \bar{Y}, \bar{t}$ defined by
\begin{equation}
\label{eqinva}
\bar{t} = \alpha t \,, \quad \bar{X} = \beta X \,, \quad \bar{Y} = \beta Y \,, 
\end{equation}
where $\alpha, \beta$ are real constants to be specified. In terms of the new variables the system in Eq. (\ref{e5}) becomes
\begin{equation}
\frac{d^2 \bar{X}}{d\bar{t}^2} = \frac{\alpha^2/\beta^4}{\bar{X}^2 \bar{Y}}  \,, \quad \frac{d^2 \bar{Y}}{d\bar{t}^2} = \frac{\alpha^2/\beta^4}{\bar{X} \bar{Y}^2} \,,
\end{equation}
so that form-invariance is achieved provided $\alpha = \beta^2$. In this case, setting $\beta = 1 + \epsilon$, where $\epsilon$ is an infinitesimal parameter, the near to the identity version of the transformation (\ref{eqinva}) becomes 
\begin{equation}
\bar{t} = t + 2\,\epsilon\,t \,, \bar{X} = X + \epsilon X \,, \quad \bar{Y} =  Y + \epsilon Y \,, 
\end{equation}
which is a symmetry with generator given by Eq. (\ref{g}).

\section*{Acknowledgments} 
The author would like to acknowledge financial support of CNPq (Conselho Nacional de Desenvolvimento Cient\'ifico e Tecnol\'ogico), Brazil.

\end{document}